\documentclass[12pt]{article}
\usepackage{epsfig,psfrag}

\newcommand{\be}{\begin{equation}}
\newcommand{\ee}{\end{equation}}
\newcommand{\bea}{\begin{eqnarray}}
\newcommand{\eea}{\end{eqnarray}}

\def\id{\protect{{1 \kern-.28em {\rm l}}}}

\def\tQ{{\tilde Q}}
\def\F{{\cal F}}

\def\bS{{\bf S^3}}
\def\bP{{\bf P^1}}

\begin{document}

\begin{titlepage}

\begin{flushright}
UCB-PTH-02/56\\
UCLA-02-TEP-37
\end{flushright}

\begin{center}
{\LARGE Baryons, Boundaries and Matrix Models.}
\vskip1truecm
\end{center}

\centerline{Iosif Bena${}^{1}$,
Radu Roiban${}^{2}$ and Radu Tatar${}^{3}$
}

\bigskip

\centerline{~~~${}^{1}$~Department of Physics
~~~~~~~~~~~~~~~~~~~~~~~~~~~
${}^{2}$~Department of Physics~~~~~~}
\centerline{~~~~University of California
~~~~~~~~~~~~~~~~~~~~~~~~~~~~
University of California~~~~}
\centerline{~~~~~Los Angeles, CA 90095
~~~~~~~~~~~~~~~~~~~~~~~~~~
Santa Barbara, CA 93106~~~~
}
\centerline{\tt ~~~iosif@physics.ucla.edu
~~~~~
 radu@vulcan2.physics.ucsb.edu~~
}

\centerline{}
\centerline{${}^{3}$~Department of Physics}
\centerline{and}
\centerline{Theoretical Physics Group}
\centerline{Lawrence Berkeley National Laboratory}
\centerline{University of California}
\centerline{Berkeley, CA 94720}
\centerline{\tt rtatar@socrates.berkeley.edu}

\begin{abstract}

A natural extension of the Dijkgraaf-Vafa proposal is to include fields in the 
fundamental representation of the gauge group. In this paper we use field theory 
techniques to analyze gauge theories whose tree level superpotential is a generic 
polynomial in multi-trace operators constructed out of such fields. We show that 
the effective superpotential is generated by planar diagrams with at most 
one (generalized) boundary. This justifies the proposal put forward 
in \cite{BERO}.

We then proceed to extend the gauge theory-matrix model duality to include
baryonic operators. We obtain the full moduli space of vacua for an $U(N)$
theory with $N$ flavors. We also outline a program leading to a string theory
justification of the gauge theory-matrix model correspondence with fundamental matter.

\end{abstract}

\end{titlepage}


\section{Introduction}

Recently, Dijkgraaf and Vafa have 
proposed \cite{dv1,dv2,dv3} a perturbative method for computing 
the effective glueball superpotential of several classes of $N=1$ theories.
The proposal instructs one to compute the planar free energy of the matrix 
model whose potential is the tree-level superpotential of the theory of 
interest. 

Being obtained via a ``string theory route'', the original proposal 
naturally includes theories with fields transforming in the 
adjoint/bifundamental representation of the gauge group. One of the most 
natural extensions of this duality is to theories with 
fields in the fundamental representation of the gauge group. It turns out 
that this goal can be reached in a rather simple way: one needs to 
compute only the contribution to the 
free energy of the matrix model arising from Feynman diagrams with one 
boundary \cite{ACFH, BERO}. Thus, the gauge theory effective superpotential 
is constructed as
\be
W_{\it eff}(S,\Lambda)=N_cS(1-\ln{S\over \Lambda^3}) + 
N_c{\partial F_{\chi=2}\over \partial S}
+N_f F_{\chi=1}~~.
\ee

Such a construction was successfully used to compare matrix model 
predictions with known gauge theory results for theories with massive 
and massless flavors, with ${\cal N}=1$ and ${\cal N}=2$ supersymmetry 
\cite{ACFH,BERO,SUZU,janik,fehe,naculich}. 
Other interesting related work has appeared in \cite{MCGR,GOPA,ito,TACH,
Dorey,Aganagic,Ferrari,Fuji,DGKV,nacu,Berenstein}.

Of a much deeper interest has been the investigation of the structure 
underlying this duality. In \cite{grisaru} it was proven that in a 
gauge theory with adjoint fields, the computation of the superpotential 
is reproduced by the matrix computation, diagram by diagram. Using 
anomaly-based arguments, in \cite{CDSW} it was also shown that 
the gauge theory and the matrix model results are related as proposed 
by Dijkgraaf and Vafa.

In this paper we prove using field theory techniques that the 
effective superpotential of a theory with a tree level superpotential 
being a generic polynomial in multi-trace and baryonic operators is generated 
by planar diagrams with at most one (generalized) boundary. This justifies 
the proposal put forward in \cite{BERO}. 
We then proceed to extend the gauge theory-matrix model duality to include
baryonic operators \footnote{As we were preparing the manuscript, \cite{ACFH1} appeared
which has some overlap with our discussion of baryons.}. 
We obtain the full moduli space of vacua for an $SU(N)$
theory with $N$ flavors. We then outline a program for the string theory
justification of the extensions of the Dijkgraaf-Vafa duality.

\section{Field theory analysis of the effective superpotential}

In this section we analyze in detail the effective superpotential as a function of
the glueball superfield $S=Tr(W^2)$ and the various coupling constants that exist in the theory,
and show that in theories with fields transforming in the fundamental representation of 
the gauge group the superpotential is generated entirely by Feynman diagrams with a 
single boundary. The analysis is similar to the one described in \cite{CDSW}. We will first consider 
tree level superpotentials built out of traces of products of quark bilinears. This analysis proves the 
proposal put forward in \cite{BERO}, that only diagrams with one boundary contribute to the effective 
superpotential. We will then add multi-trace deformations as well as baryonic operators.

We begin by considering a theory with a fairly arbitrary, polynomial, tree level 
superpotential:
\be
W_G=\sum_{l\ge 1}g_l\mbox{Tr}[(Q{\tilde Q})^l]~~,
\label{W0}
\ee
where the trace is over the flavor indices. If no tree level superpotential is present, the global symmetry
of the theory  is  $SU(N_f)\times SU(N_f)\times \prod_{i-1}^{N_f}U(1)_i\times \prod_{i-1}^{N_f}
{\widetilde U(1)_i}\times U(1)_R$. The introduction of the superpotential above, with scalar couplings,
breaks this symmetry group to
\be
SU(N_f)\times U(1)_i\times {\widetilde U(1)_i}\times U(1)_R~~.
\label{gsg}
\ee

As in the case of a theory with adjoint fields \cite{CDSW}, these symmetries are not sufficient to completely 
determine the effective superpotential. This is easy to see by considering the following  rather unusual 
charge assignment for the various fields and couplings:
\be
\begin{array}{cccccccccccc}
                  & SU(N_f) & & U(1) & & {\widetilde{U(1)}} & & U(1)_R && \Delta \\ 
Q:              & N_f         &  & \;\; 1  &  & \;\; 0                           & & 1             & & 1 \\
\tilde{Q}: &  1             & & 0           &  & \;\; 1                           & & 1             & & 1\\
g_l:          & 1               &  & \;\;-l  &  & \;\; -l                           & & 2-2l        & & 3 - 2l\\
S:              & 1              &  & \;\; 0  &  & \;\; 0                           & & 2              & &3
\end{array}
\ee
The last column in the table above represents the engineering dimension of the corresponding field.

A simple counting argument shows that all higher loop\footnote{The one loop graphs are the 
only divergent ones; their contribution can be taken into account using anomaly arguments 
\cite{CDSW}. An attempt in this direction was made in \cite{TACH}} Feynman 
diagrams with insertions of the glueball superfield are finite. We are thus looking for a function 
which depends only on $S$ and $g_i$. Furthermore, since this function is generated perturbatively,
neither one of its arguments is allowed to appear at a fractional power. It is then not hard to see that
the basic invariant combinations are 
\be
{g_l S^{l-1}\over g_1^l}~~~~~(\forall)~~l\ge 2
\label{inv}
\ee
and they also have vanishing scaling dimension. Thus, the most general effective superpotential which 
can be generated is
\be
W_{\it eff}=S\,F({g_l S^{l-1}\over g_1^l})
\label{arbitrary}
\ee
where $F$ is an arbitrary analytic function of its arguments\footnote{The appearance of inverse powers 
of $g_1$ is allowed perturbatively, as it can be accommodate by rescaling the quark fields. Indeed, absorbing
$\sqrt{g_1}$ in both $Q$ and ${\tilde Q}$ implies that the relevant couplings to consider are
${g_l\over g_1^l}$.}. A generic term in the  series expansion of $W_{\it eff}$ is
\be
c_{l_1\dots l_n}(\prod_{i=1}^n {g_{l_i}\over g_1^{l_i}} ) S^{1+\sum_{i}(l_i-1)}~~,
\label{generic}
\ee
where the $l_i~i=1,\dots ,n$ are not necessarily distinct and $c_{l_1\dots l_n}$ are numerical coefficients. 
It is clear from the coupling constant dependence that this term is generated by a Feynman diagram 
with $n$ vertices with the corresponding couplings $g_{l_i}~i=1,\dots ,n$. The $S$-dependence of this term
also fixes the number of index loops in the diagram. Indeed, since $S$ involves a trace, it is clear that each 
index loop will generate at most one factor of the glueball superfield. This implies that the diagram will 
have at least
\be
L=1+\sum_{i=1}^n(l_i-1)
\label{i1}
\ee
index loops. As in the case of diagrams with fields transforming in the adjoint representation of the 
gauge group, it is possible to show that, for a Feynman diagram with vertices generated 
by (\ref{W0}), which has genus $g$ and $b$ boundaries, the number of gauge index loops is given by
\be
L_g=2-2g-b+\sum_{i=1}^n(l_i-1)~~.
\label{i2}
\ee
Thus, since the theory we are interested in has no fields in the adjoint representation (and thus each diagram 
has at least one boundary), the only way for the relevant Feynman diagrams to have the number of index loops required by 
(\ref{i1}) is to be planar and have exactly one boundary. This proves the proposal put forward in
\cite{BERO}.

It is easy to add multi-trace deformations to the tree level superpotential in equation (\ref{W0}). The 
coupling constants now carry several indices, counting the number of traces as well as the number of 
quark bilinears in each trace. Thus, the terms with $l$ quark bilinears are:
\be
\sum_{s_1+\dots s_m=l}g^{(l)}_{s_1\dots s_m}Tr[(Q{\tilde Q})^{s_1}]\dots Tr[(Q{\tilde Q})^{s_m}]
\ee
It is clear that the charges of $g^{(l)}_{s_1\dots s_m}$ under the symmetry group (\ref{gsg}) are the 
same as the charges of $g_l$ in the table above. Thus, the generalization of (\ref{inv}) to this case is
\be
{g^{(l)}_ {s_1\dots s_m}S^{l-1}\over g_1^l}~~~~~(\forall)~~l\ge 2~,~~(\forall)~m~~.
\label{mtrinv}
\ee
As in the previous case, the Feynman diagrams generating the term proportional to $S^{1+\sum_{i}(l_i-1)}$
must have ${1+\sum_{i}(l_i-1)}$ index gauge loops. Unfortunately, the relation between the type of vertices 
used in the diagram, the number of gauge index loops and the number of boundaries
becomes more complicated than (\ref{i2}). It is however not hard to see that the number of index loops 
remains unchanged if one replaces a single-trace vertex with a multiple-trace vertex while keeping the 
genus unchanged. 

\begin{figure}[ht]
\begin{center}
\epsfig{file=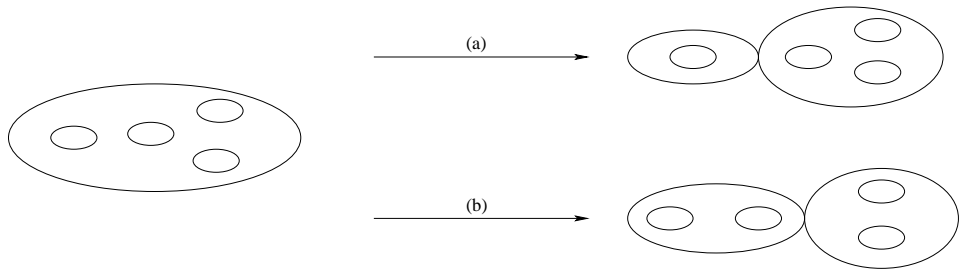,height=3.5cm}
\caption{Degenerations of the disk.\label{pinch}}
\end{center}
\end{figure}

Graphically, this corresponds to a degeneration of a disk by pinching its boundary, as 
is shown in figure \ref{pinch}. On the left-hand-side of that figure we show a vacuum diagram 
constructed out of a 4-point and a 6-point single-trace vertices. The outer line represents the boundary (corresponding 
to a trace over flavor indices) while the inner lines represent gauge index loops. On the right-hand-side we show 
two possible degenerations of this disk. Following route $(a)$, we replace the single-trace 4-point vertex by 
a double-trace one. Following route $(b)$, we  replace the single-trace 6-point vertex by a double-trace one. 
Since these surfaces lie at the boundary of the moduli space of the disk with holes,  it is clear from this 
perspective that the latter diagrams should contribute on the same footing as the former ones. With 
this definition of  boundary components, it follows that equation (\ref{i2}) still applies and the diagrams
contributing to the superpotential have exactly one boundary.

Let us now turn to including baryonic operators
\be
B=\det Q~~~~~~~~{\tilde B}=\det{\tilde  Q}
\label{B0}
\ee
if $N_f=N_c$ and
\bea
&&B^{i}=
\epsilon_{\alpha_1\dots\epsilon_{N_c}}Q_{i_1}{}^{\alpha_1}\dots Q_{i_{N_c}}{}^{\alpha_{N_c}}
\epsilon^{i_1\dots i_{N_c}\dots i}\nonumber\\
&&{\tilde B}_{i}=
\epsilon^{\alpha_1\dots\epsilon_{N_c}}{\tilde Q}^{i_1}{}_{\alpha_1}\dots Q^{i_{N_c}}{}_{\alpha_{N_c}}
\epsilon_{i_1\dots i_{N_c}\dots i}~~
\label{B1}
\eea
if $N_f=N_c+1$.
We will repeat the analysis in the beginning of this section and construct the basic invariants which can 
be built out of these operators. We will deform the initial theory only by terms linear in the baryons. More 
general deformations can be treated similarly. The charges of the baryon sources are then:
\be
\begin{array}{cccccccccccc}
                  & SU(N_f) & & U(1) & & {\widetilde{U(1)}} & &                 U(1)_R && \Delta \\ 
b:              & 1               &  & \;\; -N_c  &  & \;\; 0                               & & 2-N_c             & & 3-N_c\\
\tilde{b}: &  1              & & 0                &  & \;\; -N_c                           & & 2-N_c             & & 3-N_c
\end{array}
\ee
It is easy to see that the basic invariants are
\be
I_{l,n}={g_l\over g_1^l} \left[{b\over g_1^{N_c/2}}{{\tilde b}\over g_1^{N_c/2}}\right]^n S^{(l-1) + n (N_c-2)}
\ee
where $n$ is an arbitrary number. Thus, the superpotential is a series expansion in $I_{l, n}$.  
We can then immediately see the Feynman diagram origin of a generic term in this series:
\be
S\left[\prod _i{g_{l_i}\over g_1^{l_i}} S^{l-1}\right]
\left[{b\over g_1^{N_c/2}}{{\tilde b}\over g_1^{N_c/2}} S^{ N_c-2}\right]^{n_i}~~.
\label{gen2}
\ee
If all $n_i$ vanish, this term matches equation (\ref{generic}) and each color index loop contributes one power 
of $S$. Consider then adding a baryon and anti-baryon to the  diagram giving this contribution. To produce
a 1PI diagram we need to break one of the initial color index loops and connect it to one $Q$ and one ${\tilde Q}$
from the baryons. Thus, we find that the number of index loops increases with $N_c-2$ if the quark lines are
connected in a planar way.  Adding more baryon-anti-baryon pairs follows the same pattern and recovers the 
second factor in equation (\ref{gen2}). 

We would like to point out that the notion of planarity we use here is slightly more general than usual.
Due to the presence of the Levi-Civita symbols at the baryon vertices one is allowed to use their symmetry properties
to reorganize the internal lines of a diagram.

With hindsight, this result is not surprising in light of the discussion concerning multi-trace deformations 
of the initial theory. Since baryon-anti-baryon pairs can be re-expressed as sums of multi-trace operators, 
it follows that the counting of powers of $S$ for correlation functions involving insertions of such pairs
is identical to the counting   of powers of $S$ for correlation functions involving insertions of multi-trace 
operators. There we saw that each color index loop gave a single $S$ and the diagram was required to be 
planar and have a single boundary in the generalized sense described above.

We have thus proven that in theories with a tree-level superpotential containing multi-trace operators
and/or baryon operators, the effective superpotential as a function of the glueball superfield is generated
by planar Feynman diagrams with a single boundary in the generalized sense defined above.

\section{On the dynamics of massless fields \label{janikrev}}

One of the most puzzling aspects of the Dijkgraaf-Vafa proposal relating a gauge theory with a matrix model
is the fact that nonperturbative physics in the former can be obtained by doing perturbative 
computations in the latter.

To relate the two theories, we are instructed to first compute 
the free energy of the matrix model and construct from it the gauge theory effective 
superpotential. As it is the case with all dualities, there exists a dictionary between 
objects on both sides, but these objects are not the same. One can only compare correlation 
functions involving these objects, but not the objects themselves.
Therefore, to make contact between the two theories, one needs to integrate 
out all fields and to compare the effective superpotentials as (complicated) functions 
of the coefficients of the tree level superpotential and the dynamically generated scale.

It seems therefore that the only way one can compute gauge theory superpotentials is to 
use the matrix model to compute the effective superpotential in the ``deepest infrared''
in the Wilsonian sense, and then to integrate in all these fields. This seems to limit 
the use of matrix models in describing theories which one cannot obtain by  integrating in 
fields. 

This procedure of reconstructing the gauge theory superpotential out of the matrix model results
may seem somewhat cumbersome. Recently, for theories with tree-level superpotential expressed 
only in terms of quark bilinears (mesons), a method was proposed \cite{janik} to link the matrix 
model and the gauge theory directly, bypassing the detour through the ``deepest infrared''.  The 
method consists of introducing in the matrix path integral a $\delta$ function which relates 
the matrix model quark bilinear with the gauge theory meson. While this method seems rather ad hoc, 
as it identifies two quantities which have nothing to do with each other (they are related by a dictionary, 
but they cannot, in principle, be identified), it implies that the full effective superpotential is given by the 
sum between the tree level superpotential and an Affleck-Dine-Seiberg nonperturbatively-generated 
term. The fact that this is the correct answer can be argued using standard gauge theory arguments.

It is unclear at this time how to extend this program to directly include superpotentials with terms 
depending of several types of fields (e.g. Yukawa coupling), with terms which cannot be written in
terms of quark bilinears (e.g. baryon operators in SQCD with $N_f\ge N_c$). 

This procedure is moreover not unique. In the appendix we present an alternative procedure relating 
the matrix model and the gauge theory which also gives the correct gauge theory result and moreover 
treats different representations on the same footing. Both the procedure of \cite{janik} as well as the one 
described in the appendix are best thought of as {\it shortcuts} which relate the two sides before 
integrating out the fields on either side. It would be quite interesting if one can justify them 
to be more than this. 

Let us proceed by reviewing the method proposal of \cite{janik}. The idea is 
to identify the matrix model gauge invariant quark bilinear with the field theory meson, by introducing  
a constraint for the quarks in the matrix integral. More formally, 
\be
e^{-\F_{\xi=1}} = \int{DQ_i D{\tilde Q}_i} \delta(M_{ij} - Q_i{\tilde Q}_j) 
e^{-W_{tree}(Q_i {\tilde Q}_i)}~~,
\label{delta}
\ee
where the $\delta$ function enforces a direct relation between the
high energy description (in terms of the independent fields $Q_i{}^\alpha$ and ${\tilde Q}_\alpha{}^i$) 
and the one in terms of composite fields $M_{ij}$. 

Because of this $\delta$ function, any tree level superpotential which can
be expressed as a function of the mesons automatically gets out of the
integral. The remainder of the integral gives the volume of the
constrained quark integration domain. As in the original DV prescription,
the glueball superfield is identified with the dimension of the matrix model matrices 
(at least in the case of $U(M)$ matrix model gauge group) after it was taken to be large.
Adding the usual Veneziano-Yankielowicz terms (to account for the gauge field dynamics \cite{CDSW})
and integrating out the glueball superfield one obtains the Affleck-Dine-Seiberg superpotential. 
The net result of (\ref{delta}) is to always give the meson effective potential
\be
W_{eff}=W_{ADS}+W_{tree}~~.
\label{W-janik}
\ee

For simple tree level superpotentials this is indeed what one expects to
find in gauge theory \cite{intriligator-seiberg}. However, it is not a priori clear
that for a generic tree level superpotential the ADS superpotential is the only possible 
nonperturbative contribution. If this were not the case, the procedure  (\ref{delta}) would fail.

We can outline the sketch of a gauge theoretic argument that indeed there is no nonperturbative contribution 
besides the ADS piece. Let us assume a $W_{tree}$ of the form
\be
W_{tree}= \sum_i g_{i,k} Tr[X^i]^k 
\ee
Besides the ADS superpotential, one can expect functions of the dimensionless and chargeless ratios of 
coupling constants (e.g. ${g_1 g_4 \over g_2 g_3}, {g_1 g_5 \over g_2 g_4}, {g_2 g_{17} \over g_{12} g_7},$ etc.) 
to appear multiplying $W_{ADS}$. However, these functions need to be expandable in positive 
integer powers of the coupling constants
around the origin. Moreover they need to be holomorphic in all couplings. Hence, these functions would be holomorphic 
functions which go to zero at $\infty$, and would thus be zero everywhere. Thus, the ADS superpotential will 
be the only nonperturbative contribution to the superpotential.

As pointed out in the beginning, the procedure (\ref{delta}) is an effective way to bypass integrating 
out on both sides when the potential can be expressed in terms of mesons only. However, extending it to fields in other 
representations appears challenging.

For example, in theories containing massless adjoints the number of mesons one can
construct becomes very large \cite{kutasov,murayama}. Multiple mesons can appear also in the case one 
considers theories with massless quarks in the symmetric or the  antisymmetric representations of the gauge group.
In order to capture the dynamics of these mesons, one would need to add them by hand in (\ref{delta}). 
This would further constrain the quark integration volume. As before, any superpotential depending 
on the mesons would get out of the integral, and the constrained volume would give the nonperturbative 
superpotential as a function of all these mesons.

Nevertheless, if one tries to apply this procedure to baryons there is no way to get
sensible results. Indeed, a gauge theory baryon is a combination of $N_c$ quarks. However, the 
matrix model prescription instructs us to take the matrix model gauge group to have large 
rank, unrelated to the gauge theory $N_c$. Thus, depending upon how one chooses to treat it,
the naive baryon operator either vanishes or is not gauge invariant. A matrix theory baryon should be 
composed of $M$ quarks, where $M$ is the dimension of the matrix. It is quite clear that the matrix model
baryon and the gauge theory one have nothing in common with each other.



One can actually see that baryons are a problem not only for various computational ``shortcuts'', 
but also for formulating the correspondence 
in the picture in which one first goes to the deep infrared of both theories.
One should add a term of the form $b \epsilon^{i_1,...,i_{Nc}} \epsilon_{j_1,...,j_{Nf}}
Q_{i_1}^{j_1} ...Q_{i_{Nc}}^{j_{Nf}} $ in the gauge theory and integrate
it out. Nevertheless, it is quite clear that such a term does not make any
sense in the large $M$ $U(M)$ matrix model where the quarks transform in the fundamental 
representation of $U(M)$ but their number remains fixed to $N_f$.  The next
section is devoted to solving this problem.


\section{Baryons}

In this section we try to understand how baryons enter in the gauge theory
- matrix model correspondence. As we explained above, introducing baryon
sources {\em both} in the large $M$ matrix model potential and in the gauge theory
superpotential cannot be done. Thus, one has to take a step back and
consider the motivation behind the DV proposal.

The proposal was first formulated via a string theory route. Unfortunately,
understanding the role of fundamentals from this standpoint is rather
challenging. In the last section of this paper we will return to the question 
of how one can
understand  theories with fields in the adjoint and fundamental 
representations via 
geometrical transition. 

The purely gauge theoretical motivation of this duality was
formulated in \cite{grisaru}, and in a nutshell can be summarized
as: ``planar diagrams in theories with adjoints compute holomorphic
quantities.''

If one adopts this outlook, the matrix model is merely a tool to
systematically compute planar diagrams. Phrased in this way, the
``dictionary" between the two sides does nothing but select the planar
diagrams in the gauge theory. Of course, it is still amazing that planar
diagrams compute quantities which one normally obtains nonperturbatively,
but the ``miracle'' now pertains only to the gauge theory side.

The extension to theories with fundamentals and no baryons was conjectured in 
\cite{BERO}
to be: ``planar diagrams and diagrams with one quark boundary compute
holomorphic quantities''. In section 2  we gave a gauge 
theoretical proof, along the lines of \cite{CDSW}, that indeed only 
diagrams with one boundary contribute to the low energy effective 
superpotential.
As a side remark, we should note that there is still no 
string theory motivation for this extension.

In the same spirit as \cite{grisaru}, the matrix model with quarks and the
correct dictionary between the two sides \cite{ACFH,BERO} should be
thought of only as a tool to systematize the gauge theory computation of
these diagrams with boundaries.


If one introduces now diagrams with baryons, it is clear that they
are quite different from diagrams with quark loops. Indeed, the quark
lines are not closed, but rather go from the baryons to the
antibaryons. Nevertheless, we have shown in section 2 that a pair 
of baryons can be thought of as a multi-trace operator. We have also introduced 
a generalized notion of boundary for this kind of operators, and we have 
proven that only planar Feynman
diagrams with one generalized boundary can generate the effective 
superpotential.

The aim is therefore to analyze in a systematic way these planar diagrams.
Unfortunately, a large $M$ matrix model appears not to be any more the
right ``tool'' for selecting these diagrams. Baryon terms of the form $b
\epsilon^{i_1,...,i_{Nc}} \epsilon_{j_1,...,j_{Nf}} Q_{i_1}^{j_1}
...Q_{i_{Nc}}^{j_{Nf}} $
cannot be introduced in the Matrix Lagrangian, because the matrix quarks
are $1 \times M$ matrices, and the combinations above are not matrix
scalars.

Thus, the only types of gauge theory superpotentials one can insert in 
the large $M$  Matrix Lagrangian are the ones which can be expressed in terms
of color singlet quark bilinears. As we have seen in the previous section, in 
the this case the shortcut of \cite{janik}
can be applied, and one simply obtains the sum of the ADS superpotential and 
the tree
level one. 

Since the large $M$ matrix model cannot be used as a tool to select planar 
Feynman diagrams when baryons 
are present, the only way to count these diagrams is to select them by hand in 
a matrix 
model whose gauge group $U(M)$ has the same rank as the gauge theory gauge 
group $U(N_c)$.  
We stress that this complication is necessary only when baryons (or more 
generally operators which exist only 
for fixed values of the $N_c$) are present.

By identifying again $g^2 M$ with $S$, one obtains an effective 
superpotential $W_{eff}(S,m_i,b,\tilde b ...)$.
Counting these diagrams is quite nontrivial, especially for baryons
with many quarks, or for theories containing other terms besides quark
masses and baryons. 

If one integrates $S$ out of the matrix result
$W_{eff}(S,m_i,b,\tilde b ...)$, and integrates in the quarks one should
obtain the full gauge theory result.

It is also possible to see that, due to charge conservation, a vacuum diagram 
contains an equal number of baryons and anti-baryons. The product of a 
baryon 
and an antibaryon can be reexpressed as a multiple trace operator, by using 
the fact that a 
product of $\epsilon$ tensors can be expressed as a sum of products of 
Kronecker $\delta $ 
functions. Therefore, any nonzero vacuum Feynman diagram containing baryons can be 
written as a diagram 
containing multi-trace operators, and as such the large $M$ limit gives 
sensible results.

Alternatively, one can integrate out a quark linking the two baryon, and 
obtain an 
effective di-baryon operator which again makes sense in the large $M$ limit.
 
It is not hard to see that this discussion combined with the fact that 
multi-trace tree level 
potentials in the gauge theory only receive $W_{ADS}$ as a nonperturbative 
correction extend the correspondence to theories with baryons.

In the next section we will compute the effective superpotential for an  
$U(2)$ gauge theory
with 2 flavors, and show that one recovers the expected gauge theory result. We then 
use the shortcut (\ref{delta}) and recover the quantum moduli space of $SU(N)$ SQCD 
with $N$ flavors.


\section{Effective superpotentials for mesons and baryons from Matrix Models}

After discussing the mesons and the baryons, we now turn to calculating 
the gauge theory superpotential from matrix model. We start with a brief review of the
field theory results for theories where ${\cal N} = 2$ SUSY is broken to  ${\cal N} = 1$.

\subsection{Field Theory results}

Consider\footnote{Unlike the rest of the paper, in this section we consider the gauge 
group to be  $SU(N_c)$ rather than $U(N_c)$. We do this to conform with the papers referred to in this section.
All relevant results also hold for  $U(N_c)$.} ${\cal N} = 1$ $SU(N_c)$
theories with $N_f$ quarks in the fundamental representation,
obtained from ${\cal N} = 2$ $SU(N_c)$ by turning on a mass term for the adjoint chiral multiplet:
\bea
W = \sqrt{2} \mbox{Tr}(\tilde{Q} \phi Q) + \mu \mbox{Tr} \phi^2
\eea
To discuss the moduli space of this theory, we distinguish between the case when the
mass of the adjoint field is small (the adjoint field remains as a field in the 
${\cal N} = 1$ theory) and the case when the mass of the adjoint field is large and the
adjoint is integrated out and we obtain  ${\cal N} = 1$ supersymmetric QCD with the
dynamical scale:
\bea
\Lambda_{{\cal N} = 1}^{3 N_c - N_f} = \mu^{N_c} \Lambda_{{\cal N} = 2}^{2 N_c - N_f}
\eea
After integrating out the adjoint field, one obtains:
\bea
\label{massadj}
W = \frac{1}{2 \mu} 
\left[
\mbox{Tr}[(Q \tilde{Q} )(Q \tilde{Q})] 
- \frac{1}{N_c}\mbox{Tr}(Q \tilde{Q})  \mbox{Tr}(Q \tilde{Q})
\right]
\eea
For $\mu \rightarrow \infty$, this superpotential approaches zero and we then obtain
${\cal N} = 1$ SQCD. 

In \cite{argyres,hoo}, a detailed discussion was presented, relating the vacua of 
${\cal N} = 2$ SUSY and ${\cal N} = 1$ SUSY theories. Without going into details, we
recall now the main results regarding the ${\cal N} = 1$ vacua:

$\bullet$ for $N_f \le N_c - 1$, the strong gauge dynamics corrects the tree level superpotential 
by the addition of the ADS superpotential (\ref{massadj}).
The moduli space is the extrema of this superpotential and, at finite values for the mass of the adjoint field, 
it has been shown to have 
2 types of solutions \cite{hoo}, one when all the diagonal entries of $M = Q \tilde{Q}$ are equal to
\bea
M_{ii} = (\frac{N_c}{N_c - N_f})^{\frac{N_c - N_f}{2N_c - N_f}}\mu\Lambda
\eea
and the other one when there are two different diagonal entries (there are only two possible
values for the diagonal entries). The presence of this two choices remains valid if we add
masses for the quarks and integrate them out. The first case has been discussed in matrix model in 
\cite{ACFH,BERO}. The second one can be treated similarly.

In a pure $N=1$ SQCD, with vanishing tree level superpotential, all these vacua run away to infinity.

$\bullet$ for $N_f = N_c$, the situations changes because the classical moduli space of vacua is 
not only parametrized by mesons, but also by baryons. In the case  $N_f = N_c$, the classical moduli 
space of vacua is modified quantum mechanically by
\be
\label{cons}
\mbox{det} M - B \tilde{B} = \Lambda^{2 N_c}
\ee
In this case there is no nonperturbatively-generated correction to the effective superpotential; 
the constraint (\ref{cons}) is implemented 
with a Lagrange multiplier $X$. The exact effective superpotential is then
\bea
W_{eff} = W = X(\mbox{det} M - B \tilde{B} -\Lambda^{2 N_c} )+ 
\frac{1}{2 \mu} \left[\mbox{Tr}(M^2) - \frac{1}{N_c}
\mbox{Tr}(M)  \mbox{Tr}(M)\right] ~~.
\eea

As before, $M$ can again have at most two different eigenvalues; for $X = 0$ we return to the
situation discussed before, for  $X \ne 0$ one has new branches which did not exist for 
$N_f \le N_c - 1$:
\bea
M_{ii} = \Lambda^{2 N_c}
\eea
for $B \tilde{B} = 0$ or
\bea
M_{ii}^{N_c} - B \tilde{B} = \Lambda^{2 N_c}
\eea
for $B \tilde{B} \ne 0$

$\bullet$ for $N_f \ge N_c+1$, the classical moduli space is not modified, but the number of baryons 
increases. In this range, there is a dual magnetic description based on the gauge group
$SU(N_f - N_c)$ with $N_f$ fundamental flavors and gauge invariant fields $M$, the Seiberg-dual 
theory \cite{seibdual}. Discussions on the realization of  Seiberg's duality for mass deformed theories 
in the context of gauge theory-matrix models appeared in \cite{fehe}, but it is unclear how to obtain 
a duality for massless theories. 

\subsection{Quantum Moduli Spaces from Matrix Models}

As explained earlier, the fields appearing 
in the matrix model are not directly related to the dynamical gauge theory fields.
Thus, one cannot directly identify the matrix model effective action in which only 
some fields were integrated out with the gauge theory superpotential.
Instead, to find the gauge theory effective superpotential 
using the matrix model, we take the following steps:

1) First we deform the gauge theory  by including mass terms for all fields as well as sources for 
the operators describing the directions in the moduli space we are interested in. In the case of 
mesons these two coincide, as the mass of a quark acts as a source for a meson as well.

2) We compute the planar free energy of the matrix model associated to this deformed 
theory.

3) We equate the resulting function with the gauge theory effective superpotential
at energy scales below the lightest mass present in the deformed theory. The 't~Hooft 
coupling of the matrix model is identified with the gauge theory glueball superfield.
Furthermore, the Veneziano-Yankielowicz superpotential is added to take into account the gauge 
dynamics.

4) We integrate in gauge theory fields by appropriately performing Legendre 
transforms

5) We restore the masses of the original fields, by appropriately taking the 
deformation parameters to zero.

To illustrate this program let us discuss in detail  an $SU(2)$ gauge theory with 
2 flavors, $Q_i$ and ${\tilde Q}_i$ with $i=1,\, 2$. This means that the adjoint field 
(appearing in the discussion in the previous subsection) has been integrated out and we
work within ${\cal N} = 1$ SQCD. We return at the end of this section to the case
when the adjoint field has finite mass. 
We start with a massless theory 
whose tree level superpotential is given by the baryon and antibaryon sources:
\be
W_{\it tree}=\int d^2\theta \, \left[
   {           b} {           Q}_1{}^\alpha {            Q}_2{}^\beta\epsilon_{\alpha\beta} 
+{\tilde b} {\tilde Q}_1{}^\alpha {\tilde Q}_2{}^\beta\epsilon_{\alpha\beta} 
\right]+hc
\ee
The first step then instructs us to define this theory as the massless limit 
of the theory with the superpotential above deformed by mass terms:
\be
W_{\it deff}=\int d^2\theta \, \left[mQ_1{}^\alpha {\tilde Q}_\alpha^1
+mQ_2{}^\alpha {\tilde Q}_\alpha^2\right]+hc
\ee 

As discussed earlier, it is clear that $W_{\it tree}$ cannot be used as part of the 
potential of a large $N$ matrix model (or any other theory for that mater), because 
it manifestly vanishes. Thus, the way to proceed is to consider an $SU(2)$ matrix 
model and compute only the planar free energy. Thus, we have the following matrix 
model potential:
\be
W_M=(Q_1{}^\alpha,Q_{2}{}^\alpha,{\tilde Q}_\alpha{}^1,{\tilde Q}_\alpha{}^2)
\pmatrix{ 0& b \epsilon_{\alpha\beta}&m & 0 \cr
                    b\epsilon_{\alpha\beta}& 0 & 0 & m \cr
                    m& 0& 0 & {\tilde b} \epsilon^{\alpha\beta}\cr
                    0  & m& {\tilde b} \epsilon^{\alpha\beta}& 0\cr}
\pmatrix{Q_1{}^\beta\cr Q_{2}{}^\beta\cr {\tilde Q}_\beta{}^1\cr {\tilde Q}_\beta{}^2\cr}
\equiv {\hat Q}^T{\hat M}{\hat Q}
\ee

Since this potential is quadratic in the dynamical fields $Q$ and ${\tilde Q}$, the free 
energy is trivial to compute. The result is:
\be
{\cal F}=-\ln Z = \ln \left[{1\over \Lambda^{2\times 2}}det {\hat M} \right]^{1/2}= 
2\times 2 \times{1\over 2} \ln({m^2 - b{\tilde b}\over \Lambda^2})
\ee
In the expression above the first factor of $2$ is due to having two flavors while the 
second factor of $2$ is due to having two colors. This latter numerical coefficient combines
with the coupling constant into the glueball superfield. Therefore the gauge theory 
superpotential is:
\be
W_G=N_cS\left[1-\ln{S\over \Lambda^3}\right] + {1\over 2}N_f S \ln({- b{\tilde b}+m^2
\over \Lambda^2})
\ee
where in the case at hand $N_f=N_c=2$ but we wrote them explicitly to 
stress the functional dependence, and we inserted $\Lambda$ for dimensional reasons.

Now we perform the Legendre transform required for integrating in the quarks in 
the form of the gauge theory mesons $X_i=Q_i{\tilde Q}^i$ and baryons 
$B=Q_1{}^\alpha Q_2^\beta\epsilon_{\alpha\beta}$ and 
${\tilde B}={\tilde Q}_1{}^\alpha{\tilde Q}_2^\beta\epsilon_{\alpha\beta}$. This 
implies that we deform  $W_G$  by 
\be
W_L=- m N_f X - b B - {\tilde b}{\tilde B}
\ee
and integrate out the sources $m$, $b$ and ${\tilde b}$. Here we also considered all 
mesons to be identical. Eliminating the sources by their 
equations of motion and using the equality between $N_c$ and $N_f$
we find:
\be
W_{\it eff}(X, B, {\tilde B}, S)=2S\ln{\Lambda^4\over - B{\tilde B} + X^2}~~.
\ee
Upon eliminating $S$ by its equation of motion we find the correct gauge theory 
superpotential
\be
W_{\it eff}=0
\ee
together with the quantum corrected moduli space:
\be
\det X - B{\tilde B}=\Lambda^4
\ee
where we wrote $X^2$ in the $SU(2)\times SU(2)$ invariant form. This reproduces the 
gauge theory result.

\subsection{The Effective Potential from Integrating Out the Massive Quarks}

Let us now consider the general case of an $SU(N_c)$ theory with $N_f=N_c$  flavors and
discuss the relation between the matrix model and gauge theory. Since we are interested in incorporating baryonic operators 
in the matrix model, we deform the gauge theory action by source terms for them. Then, the idea we will pursue 
is the following. Using matrix model reasoning we reduce the matrix model with $N_f=N_c$ flavors  to one with 
$N_f=N_c-1$ flavors. Then, for this new matrix model we construct the corresponding gauge theory superpotential. Then we integrate 
in, as gauge theory quark, the quark which was integrated out as a matrix model one as well as the baryons. To make
a long story short, we will recover the known gauge theory result, i.e. a vanishing effective superpotential and 
a quantum-deformed 
moduli space.

To begin with, the deformed tree level superpotential is:
\be
W_{\it def}=mQ_1^\alpha{\tilde Q}_{1\alpha}+bB+{\tilde b}{\tilde B}
\ee
where the baryon operators $B$ and ${\tilde B}$ are defined in equation (\ref{B0}).  Since $B$ and ${\tilde B}$ 
are antisymmetric products of the $N_f$ quarks, we can split them into products of the massive quark and the remaining 
massless quarks:
\bea
B = Q_{1 \alpha} \hat{B}^{\alpha}
\eea
where $\hat{B}{}^{\alpha}$ is product of $Q_2,\cdots,Q_{N_c}$. The same applies for the antibaryon:
\bea
\tilde{B} = \tilde{Q}_{1 \alpha} \hat{\tilde{B}}{}^{\alpha}
\eea
With these slight adjustments, the superpotential reads
\bea
W_{\it def}=m Q_1 \tilde{Q}_{1} + b  Q_{1 \alpha} \hat{B}{}^{\alpha} +
\tilde{b} \tilde{Q}_{1 \alpha} \hat{\tilde{B}}{}^{\alpha}
\eea
Since the massive quarks appear only linearly and quadratically, it is easy to compute the matrix model effective 
action by integrating out $Q_1$ and ${\tilde Q}_1$ as matrix model fields. The computation is entirely 
straightforward and yields
\bea
W_{\it eff}^{massive}=S \mbox{ln}{m\over \Lambda} -  
\frac{b \tilde{b}}{4m} 
(\hat{B}{}^{\alpha} \hat{\tilde{B}}{}_{\alpha})
\eea
where we identified the size of matrices in the matrix model with the gauge theory glueball superfield.

Because $\hat{B}^{\alpha}$ and $\hat{\tilde{B}}{}^{\alpha}$ are products of $N_{c}-1$ flavors, 
\bea
\hat{B}{}^{\alpha} \hat{\tilde{B}}{}_{\alpha} = \mbox{det}(M_{N_f=N_c-1})~~.
\eea

To obtain the gauge theory superpotential one should also integrate over the 
massless quarks. Instead of computing the free energy with this effective potential and then 
integrating in the gauge theory quarks, we can just apply the prescription \cite{janik} which we 
reviewed in section \ref{janikrev}.  This straightforwardly gives:
\bea
W_{eff}(S, M_{N_c-1}, m, b, {\tilde b}) &=& S~(- \mbox{ln}\frac{S}{\Lambda^3} + 1)
- S~\mbox{ln} \left[\frac{\mbox{det}(M_{N_f=N_c-1})}{{\Lambda_{N_f=N_c}^{2N_c-1}}}\right]\nonumber\\
&+&S \mbox{ln}{m\over \Lambda} -  
\frac{b \tilde{b}}{4m} 
(\hat{B}{}^{\alpha} \hat{\tilde{B}}{}_{\alpha})~~.
\eea
The term proportional of the logarithm of the mass of the quark which was integrated out
combines with the initial dynamical scale $\Lambda$ to give the scale of the theory with $N_f=N_c-1$ quarks,
just as in a purely field-theoretic analysis. Then, to construct the gauge theory superpotential in terms of meson fields only,
we integrate out the glueball superfield and find the usual ADS superpotential:
\bea
W_{eff}(M_{N_c-1}) = \frac{\Lambda_{Nf=N_c-1}^{2~N_c + 1}}{\mbox{det}(M_{N_f=N_c-1})} -  
\frac{b \tilde{b}}{4m} 
(\hat{B}{}^{\alpha} \hat{\tilde{B}}{}_{\alpha})
\eea

The last step is to integrate in the gauge theory quark corresponding to the matrix model quark which was integrated out, 
and restore
the last meson as well as the two baryons. The steps are identical to those in \cite{int} and we will not repeat them here.
The conclusion of this short computation is that we recover a vanishing superpotential as well as the usual modification 
of the classical moduli space of the theory.

\section{Geometric Transitions, Matrix Models and Gauge Theories for Fundamental Quarks}

The geometric transition was first formulated in type IIA string theory from one side with D6 branes wrapped 
on the ${\bf S^3}$ cycle of a deformed conifold to another side with fluxes on the $\bP$ cycle
of a resolved conifold \cite{vafa} (see \cite{vafa1,vafa2,vafa3,vafa4} for subsequent work on type IIA). 

The mirror picture is a type IIB transition from D5 branes
wrapped on the $\bP$ of a resolved conifold to  fluxes on a ${\bf S^3}$ cycle of the deformed conifold. 
The type IIB picture was easier to generalize, by using ${\cal N} = 1$ deformations
of resolved ${\cal N} = 2$ ADE singularities \cite{civ,eot,ckv,civ1,fiol}. 

The geometrical picture was the backbone of the recent exciting developments in matrix model/gauge theories.
In \cite{dv1}, the model of \cite{civ} has been used to relate matrix model variables with
geometrical variables and with gauge theory variables. 
In the comparison between geometry, matrix models and gauge theories, what needs to be 
compared is:

$\bullet$ in field theory we consider the quantum moduli space. One of the tools to study this are the 
Seiberg-Witten curves. Going beyond the field theory, gravitational corrections to field theory 
have been discussed in 
the matrix model description \cite{klemm,dijk}.

$\bullet$ in geometry, we consider the 1-complex dimensional curve obtained by reducing the 
deformed geometry on an $S^2$ fiber.

$\bullet$ in the matrix model, we consider the equation for the resolvent of the matrix.

If we are to compare just field theory and matrix model, in \cite{dv3} has been suggested that 
the planar diagrams in matrix models give the non-perturbative corrections in the field theory (we have
discussed extensively in our paper about this). In the line of \cite{dv1}, it is interesting
to get a connection between field theory and matrix model, through 
geometrical quantities. It would be interesting to also describe matter (besides bifundamental
fields) in the geometrical setting. 
It is not yet known how to go through the transition with fundamental matter, and in the present 
section we will outline a possibility, leaving the details for a future work.

To begin with, we have to know how to include fundamental matter in the geometrical set-up. 
The fundamental matter would correspond to either additional D5 branes wrapped on additional 
non-compact cycles of the conifold or to additional D7 branes which are 
partially wrapped on the $\bP$ cycles. In order to see this, one can use
the MQCD picture for the geometrical transition \cite{dot1,dot2,dot3,dot4}. In the absence of the 
fundamental quarks, the configuration with D5 branes wrapped on $\bP$ cycles goes into a brane
configuration with D4 branes between two orthogonal NS branes \cite{hw}
(there are two orthogonal lines 
of singularity in the geometry, which become two orthogonal NS branes after the T-duality).   
In the brane configuration, the fundamental matter can be introduced as either semi-infinite
D4 branes ending on one of the NS branes or as D6 branes in between the NS branes. The D4 
branes are 
better suited to describe massive flavors and the D6 branes are better suited to describe 
massless
flavors. In the geometric picture the semi-infinite D4 branes become D5 branes on a non-compact 
$\bP$ which is distanced from the exceptional $\bP$ by a distance equal to the mass of the flavor
\cite{dot2}. We do not yet have a clear picture on the geometry side for the 
D6 brane,
but we can use the geometry/brane configuration duality in order to discuss the transition.

For massive matter, in the infrared we go to the effective theory where all the massive matter 
has been 
integrated out and the sizes of the different $\bS$ cycles on the geometric side are related to the different 
gluino condensates in field theory, which are in turn proportional to the masses of the quarks. If one of 
the masses of the quarks becomes zero, 
the size of the $\bS$ cycle becomes zero and the geometric transition fails. 
In this case, we should use the brane configuration with D6 branes instead of the one with 
semi-infinite D4 branes. 
By lifting it to MQCD and going through the transition, the moduli space of the field theory should be 
recovered from the geometry. As the transition consists on closing the interval between the two NS branes
(going from a curved M5 brane to a plane M5 brane \cite{dot1,dot2}), we expect to get a geometry described
by an equation $x y = a$ in the $(x,y)$ plane. But section 5.2 of \cite{hoo} tells us that the only case 
when $x y \ne 0$ is when the mass of the ${\cal N} =2$ adjoint field is not infinite. This is 
consistent with the fact that there is no supersymmetric vacua for the corresponding ${\cal N} =1$ theory.
In the case of infinite mass for the adjoint field, after integrating it out in the matrix theory, 
we remain with decoupled integrals over the massless fundamental quarks (with no tree level superpotential), 
and this gives the ADS superpotential which removes any supersymmetric vacua. 

In order to obtain a vacuum, we have to keep the mass of the ${\cal N} =2$ adjoint field as a finite 
quantity. In this case the matrix integral becomes harder to compute, as there are extra couplings between fields.
This is currently under investigation.

We can ask whether there exists a geometrical picture corresponding to finite mass for the 
${\cal N} =2$ adjoint field. Remember that the starting point of the geometrical 
discussion was the case when the mass of the ${\cal N} =2$  adjoint field is $\infty$. In the 
brane configuration this comes from the fact that  the NS branes are orthogonal and this
comes from the orthogonality of two lines of singularity in the geometry. As we know from brane 
configurations, the angle between the NS branes is related to the mass of the adjoint, 
therefore a finite 
mass for the adjoint field can be obtained only when the NS branes are not orthogonal which 
would imply a geometry with non-orthogonal lines of singularity.

\section{Conclusions}

In this paper we discussed several aspects of the Dijkgraaf-Vafa relation 
between matrix models and gauge theories. Our investigations focused
on the fields transforming in the fundamental representation of the gauge group.
Using field theory arguments (based on symmetries and holomorphy) we proved 
that the effective superpotential for the 
glueball superfield is generated on the gauge theory side by planar (in a slightly 
generalized sense) Feynman diagrams
with exactly one boundary. Our result holds for a tree level superpotential 
given by a generic polynomial in multi-trace operators and baryonic couplings.
It is then likely that, using techniques similar to those in \cite{grisaru},
the gauge theory computation be explicitly reduced to a matrix model computation.
A slight subtlety in the diagrammatic analysis is due to the existence of graphs 
which are not obviously planar, but can be mapped into manifestly planar ones
using the symmetries of the baryonic vertices.

We also analyzed the inclusion of baryonic sources in the matrix model computation.
This solution turned out to be rather subtle, as one cannot include the baryon operators 
in a large $N$ matrix model computation. Instead one must keep the rank of the matrix model 
gauge group equal to the rank of the gauge theory one. Then, the restriction to planar 
diagrams is not automatic any more and must be done by hand.
We explicitly performed the computations for $U(2)$ SQCD with two flavors and then 
used a ``computational shortcut'' to analyze the general case of  $U(N)$ SQCD with 
$N$ flavors. We recovered the known gauge results for the quantum moduli space.

As the geometry was a crucial tool in deriving the relation between 
matrix models and gauge theories in \cite{dv1}, it remains an interesting open
problem finding the deformations of the geometry which correspond to the addition of 
fundamental matter to the gauge theory. We have outlined a possible way to 
describe the required geometrical transition. We plan to return to this in a future work.


\bigskip\bigskip

{\bf \large Appendix}

In this appendix, using heuristic arguments, we propose a computational shortcut in the matrix 
model which yields the gauge theory effective superpotential.

The proposal implies that the effective superpotential is written as a sum of terms, each 
arising from one particular field in the theory. We will be concerned only with the 
contribution of matter fields. As in the usual DV approach, the gauge field dynamics 
is taken into account by the addition of the Veneziano-Yankielowicz superpotential 
$W_{VY}$ to the matrix model free energy.

Our proposal is that the free energy of the matrix model associated to some gauge theory
is given by the logarithm of its non-normalized partition function. Then, the gauge theory
effective superpotential is given by the (extended) DV relation:
\be
\label{efectiv}
W_{effective} = W_{VY} 
+ {\partial\F_{\chi=2} \over \partial S}+\F_{\chi=1}~~.
\ee

An immediate consequence of this proposal is that a field not appearing in the superpotential
contributes a factor of its ``integration volume'':
\be
\int D\Psi_i = {\rm Vol(\Psi_i)}~~.
\ee
As usual, we identify the 't~Hooft coupling in the matrix model, $g^2 M$,(where $M$ is
dimension of the matrices) with the glueball superfield $S$.

In the case of massless adjoints, their volume of integration is the volume of $SU(M)$ 
multiplied by that of a noncompact piece. When properly regularized the volume of the 
noncompact piece becomes an arbitrary scale.
This contribution comes in the numerator, and can cancel the ``volume of gauge group'' which 
gives the VY term. If the gauge theory was $N=2$ supersymmetric, there should not be any 
superpotential\footnote{The full analysis of the $N=2$ 
dynamics is far beyond the scope of this proposal, see \cite{naculich} for steps in that directions.} 
for the glueball superfield $S$. 
This cancellation can be achieved if one identifies 
the cutoff in the volume of the adjoint field with the dynamically generated scale, $\Lambda$.

Let us consider now the case of massless fields in the fundamental representation. Their integration volume is 
also infinite, and one can also handle it by imposing a cutoff. However, it is not {\it a priori} clear 
what this cutoff should be identified with in the gauge theory. It is nevertheless clear that one should
use an object invariant under all symmetries of the (gauge) theory which also has the right dimension.
A natural choice seems to be the determinant of the meson field:
\be
\det X = |\!\!|Q{\tilde Q}|\!\!|_{\it max}~~.
\ee

To compute the integration volume for one quark we notice that the measure $d Q~ d\tQ$ spans 
two fundamental representation; their product can be thought of as giving an 
adjoint representation and a singlet. The volume of the adjoint is finite, 
but the volume of the singlet is infinite. If we regularize this volume by 
imposing a cutoff at a ``radius'' $X$ , the total volume will 
just be volume of the adjoint multiplied by $X^{M^2}$~~ 
\footnote{This can be easily understood by analogy with the volume of a ball. 
The ball can be thought as a product of a sphere and $R^+$. The total volume 
is the product of the volume of the sphere and the radius to a power equal 
to the dimension of the space.}.

For each of the $N_f$ quarks not appearing in the superpotential, one has to consider the corresponding 
integration volume. The cutoff for each volume will be interpreted in the 
gauge theory as the meson corresponding to that quark. Due to the various global symmetries
of the theory these cutoffs should combine to form an invariant, e.g. $\det X$. 
It is not hard to see that, the net result of these considerations 
is a $W_{eff}(S,X_{ij})$ which yields the Affleck-Dine-Seiberg superpotential
upon integrating out the glueball superfield.

Before integrating out the glueball superfield, the superpotential can be justified to be
given by:
\be
W=N_c S \left[ \log \left(\Lambda^3 \over S \right) + 1\right] - N_f S \left[ \log \left(\Lambda^3 \over S \right) + 1\right] - 
S \log ({\det X_{ij}\over \Lambda^{2N_f}} ) 
\label{W}
\ee
Indeed, the first term is the contribution of the gauge field; the second term comes from the compact part of the integration 
volume of the quarks; finally, the last therm is given by the non-compact directions in the quark integration volumes, 
combined in a way compatible with the global symmetries of the theory.


The mesons  $X_i$ appear now as cutoffs (regulators) in the integration volumes, and each 
gives a contribution to $\F$ of the form $S^2 \log X_i$, as explained above. This justifies the 
$X$ dependence of (\ref{W}), as the superpotential 
is related to the first derivative of $\F$ with respect to $S$,

When $N_f=N_c$, the first two terms cancel, and the only part left is
$S \log {\det X\over \Lambda^{2N_c}}$. Integrating out $S$  gives a vanishing superpotential as well as the
constraint
\be
\Lambda^{2N_f}=\det X 
\ee
which is the correct gauge theory result in the absence of baryons.

\bigskip\bigskip

{\bf Acknowledgments}

\vskip 3mm
We thank Eric d'Hoker, Per Kraus and Hitoshi Murayama for useful discussions. 
The work of I.B. was supported by the NSF under 
Grant No. PHY00-99590. The work of R.R. was supported in part by DOE under 
Grant No. 91ER40618 and in part by the NSF under Grant No. PHY00-98395.
R.T. was supported by the DOE grant DE-AC03-76SF00098, the NSF grant PHY-0098840 
and by the Berkeley Center for Theoretical Physics.



\end{document}